\documentclass[showpacs]{revtex4}

\usepackage{graphicx}
\usepackage{amsmath}
\usepackage{epsfig}
\usepackage{amssymb}
\usepackage{amsfonts}

\makeatletter
\def\btt#1{\texttt{\@backslashchar#1}}
\DeclareRobustCommand\bblash{\btt{\@backslashchar}}
\makeatother

\newcommand{\beq}{\begin{equation}}
\newcommand{\eeq}{\end{equation}}
\newcommand{\bea}{\begin{eqnarray}}
\newcommand{\eea}{\end{eqnarray}}

\newcommand{\tr}{\mbox{Tr}}
\newcommand{\trs}{\mbox{Tr}_{sym}}
\newcommand{\half}{\frac{1}{2}}
\newcommand{\third}{\frac{1}{3}}

\newcommand{\sigmav}{{\vec \sigma}}

\newcommand{\lambdav}{{\vec \lambda}}
\newcommand{\nv}{{\vec n}}
\newcommand{\mv}{{\vec m}}
\newcommand{\ndn}{\vec{n}\cdot\vec{n}}
\newcommand{\nsn}{\vec{n}\star\vec{n}}
\newcommand{\nsndn}{\nsn\cdot\nv}

\def\notpa{\hbox{{$\partial$}\kern-.54em\hbox{\raisenot}}}
\def\notA{\hbox{{$A$}\kern-.54em\hbox{\raisenot}}}

\def\Bid{{\mathchoice {\rm {1\mskip-4.5mu l}} {\rm
{1\mskip-4.5mu l}} {\rm {1\mskip-3.8mu l}} {\rm {1\mskip-4.3mu l}}}}

\begin{document}

\title{Characterization of the Positivity of the Density Matrix in 
Terms of the Coherence Vector Representation}

\author{Mark S. Byrd}
\author{Navin Khaneja}

\date{\today}

\affiliation{Harvard University, Division of Engineering and Applied 
Science, 33 Oxford Street, Cambridge, Massachusetts 02138}

\begin{abstract}
A parameterization of the density operator, a 
coherence vector representation, which uses a basis of orthogonal, 
traceless, Hermitian matrices is discussed.  Using this 
parameterization we find the region of permissible vectors which 
represent a density operator.  The inequalities which 
specify the region are shown to involve the Casimir invariants 
of the group.  In particular cases, this allows the 
determination of degeneracies in the spectrum of the operator.  
The identification of the Casimir 
invariants also provides a method of constructing quantities which 
are invariant under {\it local} unitary operations.  Several examples are 
given which illustrate the constraints provided by the positivity 
requirements and the utility of the coherence vector parameterization.  

\end{abstract}


\maketitle



\section{Introduction}

The density operator must satisfy three important requirements.  1) It 
must be Hermitian.  2) The trace of the density operator, when 
appropriately normalized, must be one.  3) It must be positive 
semi-definite.  The third of these requirements has been 
found to be vital in quantum information theory, and in quantum 
mechanics itself \cite{Kraus:83}.  Perhaps 
the most important place this has arisen is in the identification of 
positive and completely positive maps which can be used to 
identify entangled states \cite{Peres,Horodeckis} and to 
classify quantum channels 
(\cite{Ruskai:03,Horodecki/Shor/Ruskai} and references therein).  
For both of these problems, but in particular the latter, a 
parameterization of the density operator is often useful.  
This provides an explicit way in which 
to identify when the channel is unital, trace preserving, and/or 
completely positive (see for example \cite{King/Ruskai}).  In 
addition, positivity requirements place restrictions on physically 
realizable quantum transformations \cite{Buzek:NOT}.  

Here we represent the density operator using a basis of orthogonal, 
traceless, Hermitian matrices.  This representation is the generalization 
of the Bloch or Coherence vector for two-state systems which is commonly 
used (see \cite{Mahler:book}).  While the geometry of the 
space of density operators for two-state systems is relatively simple, 
the geometry of the space of density operators for higher dimensional 
systems is considerably 
more complicated.  The positivity (or more precisely, positive 
semi-definiteness) conditions are therefore more 
difficult to express succinctly for higher dimensional systems.  
The inequalities given in this 
paper give necessary and sufficient conditions for a Hermitian 
operator to be positive semidefinite.  

This set of inequalities can be expressed in terms of 
a distinguished set of unitary invariants, the Casimir invariants.  
This is a particularly notable relationship since the Casimir 
invariants are associated with the ``good'' quantum numbers of a 
quantum system \cite{Biedenharn:63} and thus have direct physical 
interpretation.  They specify the set of quantities which are invariant 
under a given set of unitary transformations.  This has found 
many important applications for modelling of physical systems, 
and more recently, in quantum control of 
spin systems \cite{Navin:03}.  In addition, the Casimir 
invariants and positivity requirements are expressed in terms 
of the coefficients of the characteristic polynomials.  These 
coefficients, and their ratios, were found to be entanglement 
monotones \cite{Barnum:01}.  Entanglement monotones 
could provide some insight into the problem of finding 
suitable entanglement measures since they satisfy an important 
requirement of such measures; they do not increase, on average, 
under local operations and classical communication \cite{Vidal:em}.

This paper can be divided into three main parts (excluding the 
Introduction and Conclusion).  The first part 
gives the generalized coherence vector representation of the density 
operator and the Casimir invariants in terms of the coherence vector.  
The second part gives positivity conditions for the 
density operator in terms of the trace invariants as well as the 
coherence vector.  The third part gives some examples of the utility 
of the structures presented in the first two parts.


\section{Coherence Vector/Casimir Invariants}

\label{sec:cohvector}

In this section we present a coherence vector representation 
for an $N$-state system with particular normalization relationships 
which differ, for example, from \cite{Mahler:book}.  This 
is the generalization of the Bloch sphere representation for 
two-state systems.  The coherence vector, in our parameterization, 
has unit magnitude for pure states and 
has magnitude strictly less than one for mixed states.  
We will then show how to construct the Casimir invariants 
of the system in this parameterization.  Using a completely 
analogous construction, we are able to provide a distinguished 
set of {\it local} unitary invariants for composite quantum systems.


\subsection{Pure States in $N$-Dimensions}

\label{purestategeom}

Any density operator can be expanded in any basis of orthogonal, 
traceless, Hermitian matrices.  Here we adhere to the following 
conventions.  We will use the following normalization condition 
for the elements of the Lie algebra of $SU(N)$  
\beq
\tr(\lambda_i \lambda_j) = 2\delta_{ij}.
\eeq
We will also choose the following relations for commutation 
and anticommutation relations:
\beq
[\lambda_i, \lambda_j] = 2i f_{ijk}\lambda_k
\eeq
and 
\beq
\{\lambda_i, \lambda_j\} = \frac{4}{N}\delta_{ij}\Bid + 2 d_{ijk}\lambda_k,
\eeq
where the $f_{ijk}$ are the structure constants and the $d_{ijk}$ are 
the components of the totally symmetric ``$d-$tensor.''  These two 
equations may be combined more succinctly as  
\begin{equation}
\label{eq:lilj}
\lambda_i \lambda_j = \frac{2}{N}\delta_{ij} + i f_{ijk} \lambda_k 
                      + d_{ijk}\lambda_k.
\end{equation}

Using these conventions, we may express a pure state for an 
$N\times N$ density operator as
\begin{equation}
\label{npurestate}
\rho = \frac{1}{N}\left(\Bid 
          + \sqrt{\frac{N(N-1)}{2}} \; \vec{n}\cdot \vec{\lambda}\right).
\end{equation}
This representation is called a coherence vector representation with 
$\nv$ the coherence vector.  The constant is a convenient one such that for 
pure states 
\begin{equation}
\label{pscond}
\vec{n}\cdot \vec{n} = 1, \;\;\; \mbox{and} \;\;\; 
          \vec{n}\star \vec{n} = \vec{n},
\end{equation}
where the ``star'' product is defined by
\begin{equation}
(\vec{a}\star \vec{b})_k = 
                   \sqrt{\frac{N(N-1)}{2}}\;\frac{1}{N-2} \; d_{ijk} a_i b_j.
\end{equation}
This can be proved by direct computation using Eq.~(\ref{eq:lilj}).  

Orthogonal pure states, e.g., $|a_1\rangle$ and $|a_2\rangle$ with 
corresponding density operators 
$\rho_1 = (1/N)(\Bid + \vec{n}_1\cdot \vec{\lambda})$ and 
$\rho_2 = (1/N)(\Bid + \vec{n}_2\cdot \vec{\lambda})$ 
are orthogonal if  
\begin{equation}
\theta = \cos^{-1}\left(\frac{-1}{N-1}\right),
\end{equation}
where $\theta$ is defined by $\vec{n}_1\cdot \vec{n}_2 = \cos \theta$.  
Note that for $N=2$ this reduces to the well-known fact that 
for two-state systems, the orthogonal states are represented by 
antipodal points on the Bloch sphere.

The first condition in Eq.~(\ref{pscond}) implies that the 
coherence vector must have unit magnitude.  This restricts 
the set of vectors to those that lie on the surface of the 
unit sphere $S^{N-1}$.  The second condition 
restricts the set of allowable rotations to a proper subset of 
the group $SO(N^2-1)$.  The equations are non-linear and give 
a set of constraints which restrict to the manifold 
$\mathbb{C}$P$^{N-1}$ having $2N-2$ dimensions. 
The second condition is also related to the positivity of 
density operators, a fact which is discussed further below.


\subsection{Mixed States in $N$-dimensions}

\label{mixedstategeom}

The mixed state density operator in $N$-dimensions can be 
written in the same form as the pure state case: 
\begin{equation}
\label{nmixedstate}
\rho = \frac{1}{N}\left(\Bid 
          + \sqrt{\frac{N(N-1)}{2}} \; \vec{n}\cdot \vec{\lambda}\right).
\end{equation}
with $\vec{n}\cdot \vec{n} < 1$.  
However, unlike the case for a two-state system, there are 
more constraints on the coherence vector for dimensions greater than 
two for the Hermitian matrix here to represent a positive, semi-definite 
operator.  This will be given in Section \ref{sec:charpoly/pos}.


\subsection{Casimir Invariants}

The Casimir operators are invariant operators constructed 
from the Lie algebra elements.  In particular, they form 
a maximal set of algebraically independent elements of the 
center of the algebra, formed by homogeneous polynomials in 
the generators.  A very general discussion may be found in 
\cite{Fuchs/Schweigert}, and were first constructed in 
\cite{Biedenharn:63}.  General expressions for these are 
given in Appendix \ref{app:Casinvs}.  Here we note that 
the values of these operators can be determined by their relation 
to the trace invariants.  For example, 
let us consider a density matrix, $\rho$.  For all $\rho$
\beq
\tr(\rho^2) = \frac{1}{N}(1+(N-1)\ndn).
\eeq
The quantity $\ndn$ is the value of the quadratic Casimir operator 
(see Appendix \ref{app:Casinvs}), 
which we refer to as the quadratic Casimir invariant.  An example 
of the quadratic Casimir operator is the total angular momentum 
operator.  The Casimir invariants 
are unchanged by unitary transformations on the density operator.  
Similarly, 
\beq
\tr(\rho^3) \;=\;  \frac{1}{N^2}\big[1+3(N-1)\nv\cdot\nv+(N-1)(N-2) 
		(\nv\star\nv)\cdot \nv\big],
\eeq
is clearly invariant under unitary operations.  The quantity 
$\nsndn$ is the cubic Casimir invariant.  In the appendix 
we give the expressions for $\tr(\rho^n)$, $n\leq 9$.  
One may then recursively find higher order Casimir invariants 
and show that they are indeed unchanged by unitary transformations.  
The trace invariants, $\tr(\rho^n)$, here were discussed in 
\cite{Schlienz/Mahler} where some discussion of the 
local unitary invariants were given for GHZ states.


\subsection{Constructing Local Invariants}

We can now construct a set of quantities which are invariant under 
{\it local} unitary transformations.  These invariants, like the 
Casimir invariants are a {\sl distinguished} set.  
Clearly local unitary operations preserve the Casimir invariants 
of the marginal density operators.  However, in this section 
we discuss invariants associated with the correlation matrix.  

As an example, consider the quadratic Casimir invariant
\beq
c_2 = \ndn.
\eeq
A two-qubit density operator can be expressed in a tensor 
product basis as
\beq
\rho = \frac{1}{4}(\Bid\otimes \Bid + \vec{n}_A\cdot \sigmav \otimes \Bid 
        + \Bid \otimes \nv_B\cdot \sigmav + {\cal C}_{ij}\sigma_i\otimes\sigma_j),
\eeq
Note that local unitary transformations on systems $A$ and $B$, 
denoted $U_A$ and $U_B$, conserve 
$\nv_A\cdot\nv_A$ and $\nv_B\cdot\nv_B$ respectively.  This can be 
seen as follows,
\beq
\label{eq:localad}
U_A n_A^{(i)}\sigma_i U_A^\dagger = n_A^{(i)} R^j_i\sigma_j,
\eeq
where $R \in SO(3)$.  We can therefore rewrite 
\beq
n_A^{(i)} R^j_i =  m_A^{(i)},
\eeq
and note that $\nv_A\cdot\nv_A = \mv_A\cdot\mv_A$ since the transformation 
is orthogonal.  We also know that the set of all unitary transformations 
acting on the composite system will be a subset of the matrices in 
$SO(15)$.  This implies that 
\beq
\nv_A\cdot\nv_A + \nv_B\cdot\nv_B + \sum_{ij} {\cal C}_{ij} {\cal C}_{ij},
\eeq
is also a conserved quantity.  However, we may want to ask what quantites 
associated with the correlation matrix, ${\cal C}_{ij}$, are conserved 
under local unitary transformations.  The correlation matrix has 
rows and columns labeled by the indices $i$ and $j$ respectively.  
Now consider the vector formed from the elements in each.  
Examining Eq.~(\ref{eq:localad}), 
we see that the magnitude of these vectors, is conserved by $U_A$.  
Similarly, the magnitude of the vectors formed by the columns 
is conserved.  We may express these relations as,
\beq
U_A {\cal C}_{ij} \sigma_i\otimes \sigma_j U_A^\dagger 
       = {\cal C}^\prime_{lj} \sigma_l\otimes \sigma_j, 
\eeq
where ${\cal C}^\prime_{lj}\equiv R^i_l {\cal C}_{ij}$, implies  
\beq
\sum_{i}{\cal C}_{ij}{\cal C}_{ij} 
            = \sum_l{\cal C}^\prime_{lj} {\cal C}^\prime_{lj}. 
\eeq
Similarly for $U_B$ acting on the vectors formed from the columns of 
${\cal C}_{ij}$.  Therefore under local unitary transformations of the 
form $U_A \otimes U_B$, the following quantity is conserved,
\beq
\sum_{ij} {\cal C}_{ij} {\cal C}_{ij}.
\eeq

More generally, we may determine conserved quantites 
formed from the correlation matrix which are analogues 
of the Casimir invariants.  
For the cubic Casimir invariant, for example, the following 
quantity is invariant under local unitary transformations,
\beq
\sum_{ijk,lmn} d_{ijk}d_{lmn}{\cal C}_{il} {\cal C}_{jm} {\cal C}_{kn}.
\eeq
Similarly, we could construct invariants for systems of arbitrary dimension 
as well as systems with any number of subsystems.  

The number polynomial invariants under unitary 
transformations grows rather rapidly with the 
dimension of the system under consideration \cite{Grassl/etal}.  
One might suppose that only a subset is required for constucting 
entanglement measures given that, for example, the 
square of the concurrence \cite{Hill/Wootters,Wootters:98} for two qubits 
(see Section \ref{sec:Woottersex}) is constructed from only three 
quantities which are invariant under all local unitary transformations.  
Here we have given a subset of local invariants which may well be 
useful for many quantum information processing tasks.  The set of 
invariants given by Makhlin \cite{Makhlin} 
(see also \cite{Rains:poly,Grassl/etal}) to determine equivalence 
under local unitary operations is larger than 
the number of Casimir invariants, which are included as a subset, and 
are a complete set for determining the ability of two density operators 
to be transformed into one another by local unitary transformations.  
However, since the concurrence and I-concurrence \cite{Rungta:01a}
do not rely on this large set of invariants, 
one may expect, generally, the number of invariants needed for 
the construction of entanglement measures may be far less 
than the number required for other purposes, such as local 
unitary equivalence.  

We have now shown that a density operator can be parameterized in 
terms of a set of traceless, orthogonal, Hermitian matrices and 
have constructed associated invariant quantities.  Our next 
goal is to give positivity constraints for the density operators 
that determine the allowable sets of coherence vectors $\nv$.  


\section{Characteristic Polynomial/Positivity}

\label{sec:charpoly/pos}

In this section the characteristic polynomial of the density 
matrix is expressed in terms of the trace invariants and the 
Casimir invariants.  


\subsection{The Characteristic Polynomial}

\label{sec:multinom}

In this subsection we express the characteristic polynomial in several 
different ways in terms of invariants of the group.  Consider an 
$n\times n$ complex matrix $A$ of arbitrary dimension with eigenvalues 
$p_i$.  The characteristic equation for the matrix can be written as 
(for a similar expression, see \cite{Marmo/etal})
\beq
\label{charpolysymm}
\mbox{det}(A-\lambda \Bid) = 
	\lambda^n - S_1 \lambda^{n-1} + S_2\lambda^{n-2} - +... +(-1)^{n} S_n = 0,
\eeq
where the $S_k$ are the symmetric functions given by \cite{Horn/Johnson}
\beq
S_k = \sum_{1\leq i_1 \leq \cdots i_k\leq N} \prod_{j=1}^kp_{i_j}.
\eeq
These can be written in terms of $[\tr(\rho^n)]^m$ as
\beq
S_1 = \tr(A), \;\; S_2 = (1/2) [\tr(A)S_1 - \tr(A^2)],
\eeq
and 
\bea
\label{eq:symmfcns}
S_k =\!\!&(1/k)&\!\!\!\!  [\tr(A)S_{k-1} - \tr(A^2)S_{k-2} + ...       \nonumber \\
    && 	\!\!\!		+ (-1)^{n-1}\tr(A^n)S_{k-n} + ...        \nonumber \\ 
    && 	\!\!\!		+ (-1)^{k-2}\tr(A^{k-1})S_1+(-1)^{k-1}\tr(A^k)].\;\;\;\;
\eea
This can be proved using the fact that
\bea
[\mbox{Tr}(\rho)]^N \!\!&=&\!\! \left( \sum_{k=1}^M p_k\right)^N  \nonumber  \\ 
		&=& \!\!\!\!
\sum_{\{m_k\}}(N;m_1,m_2,...,m_M)p^{m_1}_1p^{m_2}_2\dots 
p^{m_M}_M,\;\;\;\;\;
\eea
where $\{m_k\}$ is a set of integers such that $\sum_{k=1}^M m_k = N$, 
and  
\beq
(N;m_1,m_2,...,m_M) = \frac{N!}{m_1!m_2!...m_M!}.
\eeq


\subsection{Positivity}

For a given set of real numbers $\{n_1,n_2, ... ,n_N\}\in \mathbb{R}^N$, 
we would like to know when the set will represent a valid 
density operator of the form Eq.~(\ref{nmixedstate}).  
It is clear that the right hand side of Eq.~(\ref{nmixedstate}) 
has trace one and is Hermitian.  
However, the positive semi-definite property is less trivial.  

{\it Theorem:}  For a Hermitian matrix 
$\rho = (1/N)(\Bid + \sqrt{(N(N-1)/2)\;}\nv\cdot \vec{\lambda})$ 
to represent a positive semi-definite operator it is necessary and 
sufficient for $S_k \geq 0$ for all $k$.  

Sketch of proof: Since the matrix $\rho$ is Hermitian, all eigenvalues 
of the operator are real.  This implies that the coefficients of the 
characteristic polynomial are real.  They are also non-negative if and 
only if the signs of the coefficients of the characteristic polynomial 
alternate.  In fact, the number of positive roots of the characteristic 
polynomial is the number of sign changes in the sequence of coefficients 
(pages 124-5,\cite{Bronshtein/Semendyayev}). $\square$


\subsubsection{Constraints on the Coherence Vector}

The set of inequalities $S_k \geq 0$ characterizes the region 
of permissible 
vectors which represent valid, i.e., positive semi-definite, 
density operators.  The first few of these conditions, given 
directly in terms of the coherence vector, are as follows.  
For a normalized $\rho$,
\beq
S_1 = \tr(\rho) = 1.
\eeq
Here we adhere to the conventions set forth in Sections 
\ref{purestategeom} and \ref{mixedstategeom}.  Using 
the symmetric parts of the traces, denote $\trs$ given in 
Appendix \ref{traceformulas},
\bea
\label{eq:s1-s4}
S_2 &=& \half[(\tr(\rho))^2 -(\tr(\rho^2))] = \frac{N-1}{2N}[1-\nv\cdot\nv],\;\;\;\;\\
S_3 &=& \frac{1}{6} \frac{(N-1)(N-2)}{N^2}\left[ 1   
	- 3\nv\cdot\nv + 2\nv\star\nv\cdot\nv \right], \;\;\;\;\\
S_4 &=& \frac{1}{24}\frac{(N-1)(N-2)(N-3)}{N^3} \nonumber \\
     && \times \biggl[ 1 - 6 \nv\cdot\nv + 8\nv\star\nv\cdot\nv
		+ \frac{3(N-1)}{(N-3)} (\nv\cdot\nv)^2  \biggr.\nonumber \\
     && \hspace{.25in}- \biggl. \frac{6(N-2)}{(N-3)} 
	\nv\star\nv\cdot\nv\star\nv\biggr].
\eea

Higher order invariants can be calculated using the material from the 
Appendices in a straightforward albeit somewhat tedious manner.  Note 
that if the two requirements for a density operator to be a pure state 
are met, $\ndn =1$ and $\nsn = \nv$, then $S_2$ through $S_4$ 
(as well as all higher $S_k$) vanish, 
indicating a characteristic polynomial with the solution, one 
non-zero eigenvalue.  The trace being one then demands that 
this eigenvalue be one.  

It is also noteworthy that two 
density operators have the same Casimir invariants {\it if and 
only if} they have the same eigenvalues.  This follows from the 
fact that two density operators have the same Casimir 
invariants {\it if and only if} they satisfy the same characteristic 
equation.  An entanglement measure based upon an entanglement 
monotone for a bipartite pure state must be a function only of the 
eigenvalues of the marginal density operators \cite{Vidal:em}.  This 
relation between Casimir invariants and eigenvalues implies that any 
entanglement measure based on an entanglement monotone may also be 
expressed as a function of the $S_k$ or Casimir invariants of the marginal 
density operator.


\subsection{Symmetric Functions and Casimir Invariants}

\label{sec:sym/cas}

The quantities appearing in the $S_k$ are combinations of the 
Casimir invariants.  This relationship is noteworthy for 
reasons other than those just stated.  Casimir invariants can 
be used to determine degeneracies in the orbits and 
emphasizes the relation to the physical system and 
Casimirs invariants are conserved quantities used as 
labels for quantum states.  To illustrate the ability of the 
Casimir invariants to provide information about the degeneracy of the 
spectrum, we will use the three-state system as an explicit example 
and then give a brief discussion of four-state systems.


\subsubsection{Casimir Invariants for a System with Three States}

\label{sec:casdeg3}

Since the eigenvalues are invariant under unitary transformations, 
we can discuss the interpretation of the Casimir invariants in terms of 
a diagonalized density operator.  In three 
dimensions a common basis for the traceless, diagonal $3\times 3$ 
Hermitian matrices are the Gell-Mann matrices \cite{Gell-Mann:64}.  In 
this basis, we denote the two linearly independent, traceless 
diagonal matrices as 
$$
\lambda_8 = \frac{1}{\sqrt{3}}
	    \left(\begin{array}{ccc}
		1 & 0 & 0 \\
		0 & 1 & 0 \\
		0 & 0 & -2 \end{array}\right), \;\;\;\;
\lambda_3 = \frac{1}{\sqrt{3}}
	    \left(\begin{array}{ccc}
		1 & 0 & 0 \\
		0 & -1 & 0 \\
		0 & 0 & 0 \end{array}\right).
$$
For a mixed state, we may write the diagonalized form as 
$$
\rho_d \equiv \left(\begin{array}{ccc}
		a_1 & 0 & 0 \\
		0 & a_2 & 0 \\
		0 & 0 & a_3 \end{array}\right),
$$
where $\sum_i a_i = 1$.  Expanding this using 
\bea
\rho_1 &=& \frac{1}{3}\left[\Bid +\frac{\sqrt{3}}{2}\left(\sqrt{3}\;\lambda_3 
	+\lambda_8\right) \right] 
	\;= \;\; \left(\begin{array}{ccc}
		1 & 0 & 0 \\
		0 & 0 & 0 \\
		0 & 0 & 0 
			\end{array}\right), \label{rhopure1}\\
\rho_2 &=& \frac{1}{3}\left[\Bid +\frac{\sqrt{3}}{2}\left(-\sqrt{3}\;\lambda_3 
	+\lambda_8\right)\right] 
	= \left(\begin{array}{ccc}
		0 & 0 & 0 \\
		0 & 1 & 0 \\
		0 & 0 & 0 
			\end{array}\right), \label{rhopure2}\\
\rho_3 &=& \frac{1}{3}\left[\Bid - \sqrt{3}\;\lambda_8\right] 
	\;\;\;\;\;\;\;\;\;\;\;\;\;\;\;\;\;\;\; 
	= \;\;\; \left(\begin{array}{ccc}
		0 & 0 & 0 \\
		0 & 0 & 0 \\
		0 & 0 & 1 
			\end{array}\right)\label{rhopure3},
\eea
yields a density operator of the form
\bea
\rho_d &=& \frac{1}{3}\Big[\Bid + 
		\sqrt{3}\Big((a_1\sqrt{3}/2 - a_2\sqrt{3}/2)\lambda_3 
		\nonumber \\
		&&\;\;\;\;\;\;\;+(a_1/2 +a_2/2 - a_3)\lambda_8\Big)\Big].
\eea
The coherence vector is given by 
$$
\vec{n} = (0,0,a_1\sqrt{3}/2 - a_2\sqrt{3}/2,0,0,0,0,a_1/2 +a_2/2 - a_3).
$$
Since this is a positive semi-definite, Hermitian matrix, 
the density operator formed by 
$\rho_m = U\rho_d U^\dagger = \third(\Bid + \sqrt{3}U\nv\cdot\lambdav U^\dagger)$ 
is also a positive semi-definite, Hermitian operator.  With the 
appropriate restrictions on the coefficients, we may parameterize 
all three-state density matrices (and a direct generalization for 
higher dimensional systems) in this way \cite{us,Byrd/Slater}.  

For three-state systems, the following two quantities are two independent 
Casimir invariants which, in terms of the coherence vector, 
are given by
\begin{equation}
\vec{n}\cdot \vec{n} = c_2,\;\;\;\;\;\;\;\;\; 
\vec{n} \star \vec{n} \cdot \vec{n} = c_3,
\end{equation}
The first is the quadratic Casimir invariant of the group and the 
second is the cubic Casimir invariant of the group 
(see also \cite{Marmo/etal,Mahler:book}).  
The generic orbits are given by \cite{Marmo/etal},
\beq
\ndn = c_2, \;\;\;\;\; \mbox{and} \;\;\;\;\; \nsndn = c_3 \neq c_2.
\eeq
The values of $c_2$ and $c_3$ are unchanged, i.e. invariant, under 
unitary transformations of the density operator.

The square of the coherence vector is 
$$
\vec{n}\cdot\vec{n} = a_1^2 + a_2^2 + a_3^2 - a_1 a_2 - a_1 a_3 - a_2 a_3 \leq 1. 
$$
We may also calculate
\bea
\vec{n}\star\vec{n}\cdot\vec{n} &=& a_1^3 + a_2^3 + a_3^3 + 6a_1a_2a_3 \nonumber \\ 
	&&- (3/2)(a_1^2 a_2 + a_2^2 a_1 + a_1^2 a_3 + a_2^2 a_3 \nonumber \\
         && \;\;\;\;\;\;\;\;\;\;\;\;\;\;+ a_3^2a_1 +a_3^2a_2).\nonumber 
\eea
Note that $-|\nv|^3 \leq \nsndn \leq |\nv|^3$ since  
$$
(\nsndn)^2-|\nv|^6 =\frac{27}{4} (a_1 - a_2)^2(a_1 - a_3)^2(a_2 - a_3)^2 \geq 0.
$$

Degenerate eigenvalues imply the following relations.  
\begin{enumerate}
\item  if $a_1 =a_2$
\beq
\ndn = (a_1-a_3)^2 \;\;\; \mbox{and} \;\;\; \nsndn = -(a_1-a_3)^3.
\eeq
\item  if $a_2 =a_3$
\beq
\ndn = (a_1-a_3)^2 \;\;\; \mbox{and} \;\;\; \nsndn = (a_1-a_3)^3.
\eeq
\item  if $a_1 =a_3$
\beq
\ndn = (a_2-a_3)^2 \;\;\; \mbox{and} \;\;\; \nsndn = (a_2-a_3)^3.
\eeq
\end{enumerate}
Therefore, 
when the two eigenvalues are degenerate, $\nsndn\propto |\nv|^3$.  
When the two degenerate eigenvalues are greater than the third, 
the quantity $\nsndn$ is negative and when they are smaller, 
$\nsndn$ is positive.  Thus by investigating the values of the 
Casimir invariants, we are able to extract information about 
degeneracies in the spectrum.  These degeneracies correspond 
to invariant subspaces since an eigenvalue subspace spanned 
by degeneracies is invariant under unitary transformations 
on that subspace \cite{us}.  We next comment briefly on the 
four-state and general cases of identifying degeneracies.


\subsubsection{Higher Dimensions}

For $N$-state systems, 
there are $N-1$ Casimir invariants.  This is 
the rank of the group of transformations, $SU(N)$ on 
the space of density operators, and corresponds to the number 
of elements in a complete set of commuting operators.  Each 
$S_k$, when expressed in terms of the coherence 
vector, will contain a term of the form 
$(\vec{n} \star)^{\times (k-2)}\vec{n} \cdot \vec{n}$ which 
is absent from $S_{j}, \;\; j<k$.  
In the previous section it was shown that a degeneracy in the 
spectrum of the density operator was manifest in the values of the 
Casimir invariants.  When a degeneracy in the spectrum 
exists, an added symmetry of the density operator under a subgroup 
of the group of all unitary transformations exists.  This will 
determine a relation between the Casimir invariants, and thus 
reduce the number of independent polynomial invariants.  

Let us discuss the example of four-state systems.  
If the density operator for a four-state system has the following 
spectrum, $(a,b,b,b)$ then the each of the four Casimir invariants are 
proportional to powers of $|\vec{n}|$ ($c_i \propto|\vec{n}|^i$).  
If the spectrum is $(a,a,b,b)$, then all Casimirs are zero except 
the quadratic.  Spectra of the form $(a,b,c,c)$, or 
non-degenerate spectra are not as 
easily idenitfied by their Casimir invariants.  
However, there exists a readily available program, 
{\it Macaulay}, which can check the independence of the invariants, 
thereby determining the degeneracies.  Of course, if the spectrum 
is completely degenerate, then all Casimirs vanish since $\nv=0$ 
for the completely degenerate case.  The 
advantage of obtaining this information through the use of invariants 
is that {\it one may not always solve directly for the eigenvalues of a 
matrix, but the Casimir invariants may still be obtained.}  

For the convenience of the reader, the Casimir invariants are given 
in terms of the Lie algebra elements in Appendix \ref{app:Casinvs}.  
In Appendix \ref{traceformulas} we give the trace formulas from 
which these can be calculated and the coefficients of the 
characteristic polynomial can be found.  

Note that a map from a density operator to a density operator 
may be expressed as an affine map,
\beq
\nv \rightarrow \nv^\prime = T\nv + \vec{t},
\eeq
where $T$ is a matrix and $\vec{t}$ is a translation.  The 
positivity of the mapping is determined by the positivity of 
the density operator formed by $\nv^\prime$ \cite{King/Ruskai}.


\section{Examples}

In this section we give the following results.  First, we show how the 
positivity of the $S_k$ restrict the coherence vector for two particularly 
interesting examples, $\nsndn=-|\nv|^3$ and inversion.  
This gives, in terms of the coherence vector, the same 
bound obtained by Rungta, et al. \cite{Rungta:01a} 
on the ability to construct a ``universal inverter.''  
Second, we show that the positivity of the density 
operator of two qubits can be determined by the positivity of $S_3$ 
and $S_4$ for the general case and for the Werner state.  Third, we 
present an alternative derivation of the three-tangle of Coffman, Kundu and 
Wootters \cite{Coffman/Kundu/Wootters} using the coherence vector 
description.  


\subsection{Inversion of the Coherence Vector}

\label{sec:Univert}

Here we show that, due to positivity requirements, the limit 
$\nsndn=-|\nv|^3$ {\sl cannot} be reached for certain $\nv$.  
This follows from the positivity requirements $S_k \geq 0$ and 
restricts the set of positive maps for the set of density 
matrices.  An example of this is the universal inverter and 
universal NOT gate.


\subsubsection{Universal Inversion}

The universal inverter and universal NOT gate 
\cite{Rungta:01a} are related to a mapping of the form 
\beq
\label{eq:uinv}
\rho \rightarrow \Bid - \rho, 
\eeq
which is positive but not completely positive.  In terms of the 
coherence vector representation, 
\beq
\rho \rightarrow \frac{1}{N}(\Bid(N-1) - c \nv\cdot \lambdav) 
		=  \frac{(N-1)}{N}\left(\Bid - 
                 \frac{c}{N-1} \nv\cdot \lambdav\right),
\eeq
where $c = \sqrt{N(N-1)/2}$.  Thus, up to an overall 
constant, the mapping corresponds to a change in sign of the 
coherence vector and a reduction of the magnitude of the coherence 
vector.


\subsubsection{Inverting the Coherence Vector}

We might ask if there exists a physical map which will properly invert 
the coherence vector.  (Inversion of the coherence vector as a 
possible generalization 
of the concurrence \cite{Hill/Wootters,Wootters:98} 
was studied by Rungta, et al. \cite{Rungta:01a}.)  This would be of the form
\beq
\rho =\frac{1}{N}(\Bid + c \nv\cdot \lambdav) \rightarrow 
		\rho =\frac{1}{N}(\Bid - c \nv\cdot \lambdav). 
\eeq
However, this is not positive.  To see this, consider 
the matrix 
\beq
\label{eq:exdmat}
\rho = \frac{1}{N}\left[\Bid +\left(\begin{array}{cc cc}
				a & 0 & \cdots & \\
				0 & a & 0 & \cdots \\
				\vdots &  & \ddots & \vdots \\
				 0 & \dots & & -(N-1)a 
				\end{array}\right)\right].
\eeq
For this matrix $\ndn = a^2$ and $\nsndn = a^3$.  This 
gives the symmetric polynomial 
\beq
S_3 \propto 1-3\ndn + 2\nsndn = 1-3a^2+2a^3.
\eeq
This function of $\nv$ is minimum when $\nsndn = -|\nv|^3 < 0$ so that 
\beq
S_3 \propto 1-3\ndn - 2|\nsndn|= 1-3a^2-2a^3.
\eeq
For this to be positive, $a\geq 1/2$ showing that for certain 
$\nv$ the limit $\nsndn = -|\nv|^3$ cannot be obtained.  This 
is unlike the case of a Hamiltonian, or general Hermitian 
matrix, where it is acceptable to have $\nsndn = -|\nv|^3$.  
For a system with three states, and no zero eigenvalues, 
$S_3$ is the non-zero determinant of the matrix.  

For higher dimensional systems the requirement that $\rho$ 
in Eq.~(\ref{eq:exdmat}) be positive corresponds to 
\beq
\frac{1}{N-1} \geq a \geq -1.
\eeq
Now if we ask for an inversion map which is positive, we seek a 
mapping of the form
\beq
\rho \rightarrow \frac{1}{N}(b\Bid - c \nv\cdot\lambdav).
\eeq
Choosing an operator of the form Eq.~(\ref{eq:exdmat}), for 
the map to be positive we require 
\beq
b \geq (1-N)a \geq (N-1).
\eeq
This is the condition found by Rungta, et al. \cite{Rungta:01a} 
for positivity and restricts inversion to a map 
of the form in Eq.~(\ref{eq:uinv}).  This 
is a condition on the positivity of the determinant which we have 
shown is $S_N$ for an $N$-state system.  


\subsubsection{Three-State Example}

For example, let us consider a three-state density matrix of the 
form 
\begin{widetext}
\beq
\rho = \left(\begin{array}{ccc}
0.15278 & 0.036084 - i0.06250 & -0.072169 + i0.12500 \\
0.036084 + i 0.06250 & 0.23611 & -0.25  \\
-0.072168 - i0.12500 & -0.25 & 0.61111
                  \end{array}\right).
\eeq
\end{widetext}
Using $n_i = (\sqrt{3}\;/2)\tr(\rho \lambda_i)$, direct calculation gives 
$$
S_3 \propto 1-3(0.666)^2 + 2(0.666)^3, 
$$
However, when $0.666 \rightarrow -0.666$ then $S_3<0$ showing that 
inversion is not a positive map for this density operator.


\subsection{Two Qubit Entanglement}

In the next subsection (\ref{Wst}) the example of the 
Werner states for two qubits is investigated.  This mixture of a 
completely mixed and singlet state is separable if 
and only if the partially 
transposed density operator is positive semidefinite according to the 
Peres-Horodecki criterion \cite{Peres,Horodeckis}.  
In this case, $S_3$ and $S_4$ determine positivity.  
This will be shown using the coherence vector representation.


\subsubsection{A Basis for Two Qubits}

\label{sec:2qbasis}

Let a basis for the Lie algebra of $SU(4)$ be given by 
\begin{equation}
\{\lambda_i\}_{i=0}^{15} = \{ \sigma_i \otimes \sigma_j\}_{i,j=0}^{3},
\end{equation}
where $\lambda_0 \equiv \Bid_4$ and $\sigma_0 \equiv \Bid_2$.  
The labels correspond in the following way, 
\begin{equation}
\label{eq:2qbasis}
\begin{array}{lcl}
\lambda_i,\;\; i = 0,1,2,3 &\;\;\;\leftrightarrow\;\;\;& \frac{1}{\sqrt{2}}
\;\sigma_i \otimes \Bid, \;\;i=0,1,2,3, 
\\
\lambda_i, \;\;i = 4,5,6&\;\;\; \leftrightarrow \;\;\;&\frac{1}{\sqrt{2}}\;
\Bid\otimes \sigma_i,\;\; i=1,2,3,
\\
\lambda_i,\;\; i = 7,8,9 &\;\;\;\leftrightarrow\;\;\;& \frac{1}{\sqrt{2}}\;
\sigma_1\otimes \sigma_i,\; i=1,2,3,
\\
\lambda_i, \;\;i = 10,11,12 &\;\;\;\leftrightarrow \;\;\;&\frac{1}{\sqrt{2}}\;
\sigma_2\otimes \sigma_i,\; i=1,2,3,\;\;\;
\\
\lambda_i,\;\; i = 13,14,15& \;\;\;\leftrightarrow \;\;\;&\frac{1}{\sqrt{2}}\;
\sigma_3\otimes \sigma_i,\;i=1,2,3.
\end{array}
\end{equation}
This forms an orthogonal basis with respect to the trace and has 
normalization given by 
\begin{equation}
\mbox{Tr}(\lambda_i\lambda_j) = 2\delta_{ij}.
\end{equation}

The non-zero, totally symmetric d-tensor components in this basis are given by:
\bea
\frac{1}{\sqrt{2}} &=& d_{1,4,7} = d_{1,5,8} = d_{1,6,9} = d_{2,4,10} 
= d_{2,5,11} = d_{2,6,12} 
                         \nonumber \\
                   &=& d_{3,4,13} = d_{3,5,14} = d_{3,6,15} = -d_{7,11,15} 
= -d_{8,12,13} \nonumber  \\ 
&=& d_{7,12,14} = -d_{9,10,14} = d_{8,10,15}= d_{9,11,13}.
\eea


\subsubsection{Werner States: A Case Study}

\label{Wst}

Under partial transpose of the first subsystem in 
the density operator, only elements $n_2, n_{10}, n_{11}, n_{12}$ change 
sign (in the given basis Subsection \ref{sec:2qbasis}).  
Therefore under the partial transpose, one may readily 
determine which elements of the products $\nsn\cdot\nv$ and 
$\nsn\cdot\nsn$ change sign.  

The inequalities $S_3 \geq 0$ and $S_4 \geq 0$ depend only on 
the non-local invariants of the system since $S_2$ does not change 
and the local invariants which have the same form of $S_2$ also do 
not change.  This shows that the negativity arises in the nonlocal 
invariants (as they should).  As noted before, the partial transpose 
is positive since it preserves local positivity, but is not completely 
positive.  Although this is a low-dimensional example and the higher 
order $S_k$ become more complicated as the $k$ increases, such an 
analysis might lead to ways (e.g. numerical and/or analytic 
searches) for identifying positive, but not completely positive 
maps which may witness entanglement.

To clarify the discussions above concerning the positivity 
of the coefficients of the characteristic polynomial, we give an 
example of the calculation for the Werner state of two qubits.  
The Werner state \index{Werner state} for two qubits is given by 
\beq
\rho_W = \frac{1-x}{4} \Bid + x S,
\eeq
where $0 \leq x \leq 1$ is real and $S$ is the singlet state 
\beq
S = \half \left(\begin{array}{cccc}
		0 & 0 & 0 & 0 \\
		0 & 1 & -1 & 0 \\
		0 & -1 & 1 & 0 \\
		0 & 0 & 0 & 0 
			\end{array}\right).
\eeq
Therefore when $x = 0$ the state is separable and when $x=1$ the state is 
maximally entangled.  We may rewrite this as 
\bea
\rho_W &=& \frac{1}{4} \Bid - \frac{x}{4}(\sigma_x \otimes \sigma_x + 
		\sigma_y \otimes \sigma_y +\sigma_z \otimes \sigma_z )
            \nonumber  \\
       &=& \left(\begin{array}{cccc}
	\frac{1-x}{4}&          0    &        0     &       0        \\
	0            & \frac{1+x}{4} & -\frac{x}{2} &       0        \\
	0            & -\frac{x}{2} & \frac{1+x}{4} &       0        \\
	0            &     0        &        0      &  \frac{1-x}{4}
           \end{array}\right).
\eea
The partial transpose \index{partial transpose}
condition (Peres-Horodeckis) \cite{Peres,Horodeckis} is equivalent 
(up to a local unitary transformation) to the inversion of the 
coherence vector, which is also known as spin flip or inversion.  
In terms of the coherence vector for the combined system, if we 
write the density operator in terms of the basis given in the 
previous section, 
\beq
\rho_W = \rho_{AB} = 
         \frac{1}{N}\left(\Bid + \sqrt{6}\; \nv \cdot \lambdav\right), 
\eeq
the partial transpose corresponds to 
$\nv_2 \rightarrow -\nv_2,$ $\nv_{10} \rightarrow -\nv_{10},$ 
$\nv_{11} \rightarrow -\nv_{11},\nv_{12} \rightarrow -\nv_{12}$.  
Calculating the coefficients of the characteristic polynomial, 
we find $S_1(\rho_{AB})$ and $S_2(\rho_{AB})$ are unchanged under 
this transformation.  However,
\bea
S_3(\rho_{AB}) &=& \left(\frac{1}{4^2}\right)(1 - 3x^2 + 2x^3) \nonumber \\
           &\rightarrow &  \left(\frac{1}{4^2}\right)(1 - 3x^2 - 2x^3),
\eea
and 
\bea
S_4(\rho_{AB}) &=& \left(\frac{1}{4^4}\right)(1 - 6x^2 + 8x^3 - 3x^4) \nonumber \\
            &\rightarrow &  \left(\frac{1}{4^4}\right)(1 - 6x^2 - 8x^3 - 3x^4).
\eea
This partial transpose condition implies that the 
density operator is separable {\it if and only if} the partially 
transposed density operator (or the spin flipped density 
operator) is positive semi-definite.  Here we see that the 
coefficients have following possibilities for sign changes.  
For $1/3 < x < 1/2$, $S_4 < 0$, $S_3 > 0$, 
and for $x > 1/2$, $S_4 <0$ and $S_3 < 0$.  However, in each case 
there is only one change in sign for an $S_k$ and therefore one 
negative eigenvalue.


\subsection{Distributed Entanglement}

\label{sec:Woottersex}

Coffman, Kundu and Wootters \cite{Coffman/Kundu/Wootters} have 
studied  ``distributed entanglement'' which concerns the 
entanglement of various subsystems of a tripartite qubit system.  
One of their main results is the description of entanglement of 
a pure state of three qubits which is not expressible in terms 
of two-qubit relations.  Here we wish to streamline their 
argument using the material presented above and thus derive 
by alternative means the ``tangle'' of three qubits.  Consider a 
pure state of three qubits for systems we label $A,B,C$.  We will 
write the density operator in a tensor product basis,
\bea
\rho_{ABC} &=& \frac{1}{8}(\Bid \otimes \Bid \otimes \Bid 
		+ \nv_A\cdot\sigmav \otimes \Bid \otimes \Bid 
		+ \Bid \otimes \nv_B\cdot\sigmav \otimes \Bid \nonumber \\
		&&+ \Bid \otimes \Bid \otimes \nv_C\cdot\sigmav 
		+ \nv_{AB}\cdot\sigmav\otimes\sigmav\otimes \Bid \nonumber \\
		&&+ \nv_{AC}\cdot\sigmav\otimes \Bid\otimes\sigmav 
		+ \nv_{BC}\cdot\Bid\otimes\sigmav\otimes\sigmav \nonumber \\
		&&+ \nv_{ABC}\cdot\sigmav\otimes\sigmav\otimes\sigmav),
\eea
where 
$
\nv_{AB} \cdot\sigmav\otimes\sigmav \equiv (n_{AB})_{ij}\sigma_i\otimes\sigma_j
$
etc.

Since $\rho_{ABC}$ represents a pure state, the marginal density matrices, 
e.g., $\rho_{AB} = \tr_C(\rho_{ABC})$ has only two non-zero eigenvalues, 
so that the square of the concurrence may be used to write 
\bea
\label{eq:ABconc}
{\cal C}_{AB}^2 &=&(\lambda_1 -\lambda_2)^2 = \lambda_1^2 +\lambda_2^2 
          -2\lambda_1 \lambda_2 \nonumber \\
		&=&\tr(\rho_{AB}\tilde{\rho}_{AB}) - 
                     2\lambda_1 \lambda_2 \leq\tr(\rho_{AB}\tilde{\rho}_{AB}),
\eea
where $\lambda_1$ and $\lambda_2$ are the square roots of the eigenvalues of 
$\rho_{AB}\tilde{\rho}_{AB}$.  The matrix $\tilde{\rho}_{AB}$ is defined by 
$\tilde{\rho}_{AB}=\sigma_y\otimes\sigma_y\rho_{AB}^*\sigma_y\otimes\sigma_y$.

At this point our argument will differ from that of 
\cite{Coffman/Kundu/Wootters}.  Since this is a pure state, the 
Schmidt decomposition can be used to choose a preferred basis for 
subsystems $AB$ and $C$.  The reduced density matrices may be 
rewritten as (using an unnormalized coherence vector) 
\beq
\rho_{AB}=\tr_C(\rho_{ABC}) = \frac{1}{4}(\Bid + \mv_{AB}\cdot\lambdav),
\eeq 
where $\mv_{AB}\equiv(\nv_A,\nv_B,\nv_{AB})$ and 
\beq
\rho_C = \tr_{AB}(\rho_{ABC}) = \half(\Bid + \nv_C\cdot\sigmav).
\eeq  
Then, by the Schmidt decomposition these two have the same 
eigenvalues.  Therefore they satisfy the same characteristic equation 
which will have only one non-trivial $S_k$ ($S_1 = \tr(\rho)=1$), 
that being 
\beq
S_2(\rho_C) = S_2(\rho_{AB}),
\eeq
which implies
\beq
\frac{1}{4}(1+\mv_{AB}\cdot\mv_{AB}) = \half(1 + \nv_C\cdot\nv_C).
\eeq
Therefore
\beq
\label{eq:sdt}
\nv_{AB}\cdot\nv_{AB} = 1 + 2 \nv_C\cdot\nv_C - \nv_A\cdot\nv_A -\nv_B\cdot\nv_B
\eeq
Noting that
\beq
\tr(\rho_{AB}\tilde{\rho}_{AB}) = \frac{1}{4}(1-\nv_A\cdot\nv_A-\nv_B\cdot\nv_B 
				+ \nv_{AB}\cdot\nv_{AB}),
\eeq
we can use Eq.~(\ref{eq:sdt}), to write
\beq
\tr(\rho_{AB}\tilde{\rho}_{AB}) = \frac{1}{2}(1-\nv_A\cdot\nv_A-\nv_B\cdot\nv_B 
				+ \nv_{C}\cdot\nv_{C}).
\eeq
This is completely equivalent to the results in Eqs.(7) and (8) of 
\cite{Coffman/Kundu/Wootters}, the latter is repeated here:
\beq
\tr(\rho_{AB}\tilde{\rho}_{AB}) = 2(\mbox{det}\rho_A + 
                    \mbox{det}\rho_B - \mbox{det}\rho_C).
\eeq
This is needed to derive the ``first main result'' 
of \cite{Coffman/Kundu/Wootters}:
\beq
{\cal C}_{AB}^2 + {\cal C}_{AC}^2 \leq 4 \;\mbox{det}\rho_A, 
\eeq
where we have used Eq.~(\ref{eq:ABconc}).

At this point, we can calculate 
$$
4\sqrt{S_2(\rho_{AB}\tilde{\rho}_{AB})} =\tau_{ABC} 
         \equiv {\cal C}_{(A)BC}^2 -{\cal C}_{AB}^2 - {\cal C}_{AC}^2.  
$$
This quantity describes the three-way entanglement of the three qubits 
and was shown in \cite{Coffman/Kundu/Wootters} to be invariant 
under the permutation of the qubits.


\section{Conclusion}

The identification of positive but not complete positive maps 
has recently become an active area of research due to the restrictions 
it places on physically realizable quantum transformations \cite{Buzek:NOT} 
and the question of entanglement of quantum systems 
\cite{Peres,Horodeckis}.  
To aid in the study of such transformations 
this paper has presented a representation of the density operator 
in terms of traceless, Hermitian, orthogonal matrices.  We then showed 
that the Casimir invariants of 
generalized coherence vector for density operator could be calculated 
directly and information about degeneracies in the spectrum of the 
operator could be obtained for some particular cases.  
It should be noted that we have given a 
representation of the density operator in bases, but the expressions 
of the Casimir invariants and symmetric functions do not depend on 
the choice of the set of traceless, Hermitian, orthogonal matrices 
in the basis.  The region of positive semi-definite density operators 
is determined by the {\it necessary and sufficient} conditions, 
$S_k \geq 0$.  The $S_k$ were expressed in terms of the coherence 
vector and Casimir invariants.  The positivity conditions given 
here not only indicate whether a density operator has all positive 
eigenvalues, but it also indicates the number of positive eigenvalues 
in terms of the number of sign changes of the sequence of 
coefficients $S_k$.  

Superoperators which map Hermitian operators 
to Hermitian operators will preserve the reality of the eigenvalues.  
Since the eigenvalues are real, the 
coefficients of the characteristic polynomial must alternate in sign if 
the eigenvalues are to be positive.  Therefore changes in the signs 
of the $S_k$ can 
indicate positivity or non-positivity of maps of the density operator.  
Given the expressions in this paper, this statement may be utilized 
directly given an affine map of the coherence vector.  

It is interesting to note that the ``measure of purity'' of a 
density operator has arisen in several contexts.  
Consider a pure state, bipartite density operator.  The 
generalized concurrence in \cite{Rungta:01a} is simply related 
to the purity of the marginal density operator.  If $\rho_A$ is the 
marginal density operator, then the concurrence is proportional 
to $S_2(\rho_A)$ which is a measure of the purity of the 
density operator.  The state $\rho_A$ is pure {\it if and 
only if} $S_2(\rho_A)$ is zero.  The state is ``less pure'' 
if this quantity is larger.  This measure of purity is also used 
in the optimal decompositions 
discussed in \cite{Audenaert/Verstraete/DeMoor}.  One might 
consider generalizations of the ``measure of purity.''  
Certainly if $S_1$ (equal to one when the matrix has unit trace) 
and $S_2$ are the only non-zero coefficients of the characteristic 
polynomial, then $S_2$ is a ``good'' measure of purity.  However, 
if $S_2$ and $S_3$ are both non-zero, then the purity should 
be measured by two quantities since pure states necessarily have 
both quantities equal to zero.  States that are closer to 
being pure are those with smaller values of these two quantities.  
Similar arguments can be made 
for the higher dimensional $S_k$.  One might then consider 
a generalization of measures of entanglement which rely on 
this modified set of ``measures of purity.''

The set of algebraic equations given by $S_k \geq 0$ give 
a set of geometric constraints on the spaces of allowable coherence 
vectors.  This may motivate further exploration of techniques from 
algebraic geometry which has already been found useful by 
Miyake \cite{Miyake} for describing pure state separability.  

Due to the generality of the arguments here and the connections 
made between Casimir invariants, algebraic geometry and positivity, 
we believe this work provides useful relations and 
insights into the structure of positive operators.  We also hope that 
it will aid in identifying positive, but not completely positive 
maps.


\appendix


\section{Casimir Invariants}

\label{app:Casinvs}

Here we give expressions for the Casimir invariants of 
a Lie group.  For a discussion see \cite{Fuchs/Schweigert}.

The {\it Killing form} $G_{ab}$ gives the {\it metric} $g_{ab}$ on the 
vector space.  This will determine the {\it quadratic Casimir} invariant 
\begin{equation}
\label{eq:quadraticcasimir}
C_2 = \sum_{a,b=1}^N g_{ab}\lambda^a \lambda^b,
\end{equation}
where $N$ is the dimension of the vector space ($N=n^2-1$ for 
$SU(n)$ groups), and $\lambda\in {\cal L}(G)$.  Note that 
$g_{ab} \propto \sum_{c,d}f^{ac}_{\;\;\;\;d}f^{bd}_{\;\;\;\;c}$ 
is an invariant, 
symmetric tensor.  To find other invariant, symmetric tensors, one forms 
\begin{gather}
\mbox{Tr}(ad_{\lambda^{a_1}} \circ ad_{\lambda^{a_2}} \circ 
\cdots \circ ad_{\lambda^{a_n}}) 
	= \;\;\;\;\;\;\;\;\;\;\;\;\;\;\;\;\;\;\;\;\;\;\;\;\;\;\;\;\;\;\;\;\;
            \nonumber  \\
\sum_{b_1,b_2,\dots,b_n =1}^N f^{a_1b_1}_{\;\;\;\;\;\;\;b_2}
f^{a_2b_2}_{\;\;\;\;\;\;\;b_3} \dots 
f^{a_{n\!-\!1}b_{n\!-\!1}}_{\;\;\;\;\;\;\;\;\;\;\;\;\;b_n}
f^{a_nb_n}_{\;\;\;\;\;\;\;\;b_1}
\end{gather}
One can express the {\it Cubic Casimir} invariant in terms of 
the totally symmetric tensor $d_{abc}$, 
\begin{equation}
\label{eq:cubiccasimir}
C_3 = \sum_{a,b,c=1}^N d_{abc} \lambda^a\lambda^b\lambda^c.
\end{equation}
Generally these higher order invariants can be expressed in terms of the 
symmetric tensor as 
\bea
\label{eq:mcasimir}
C_m &=&\sum_{\stackrel{\scriptstyle{ a_1,a_2,\dots, 
a_{m\!-\!3}}}{ b_1,b_2,\dots,b_m}}d_{a_1b_1b_2} \nonumber \\
	&&\times
d_{a_1a_2b_3}d_{a_2a_3b_4} \dots 
d_{a_{m\!-\!2}a_{m\!-\!3}b_{m\!-\!2}}d_{a_{m\!-\!3}b_{m\!-\!1}b_m} \nonumber \\
	&&\times \lambda^{b_1} \lambda^{b_2} \cdots \lambda^{b_m}
\eea
We list the first few here in order to be explicit and to enable the 
development of the pattern.  
\bea
\label{eq:45and6casimir}
C_4 \!\!&=& \!\!\!\!\!\!\sum_{a_1,b_1,b_2,b_3,b_4}\!\!\!\!\!\!
d_{a_1b_1b_2}d_{a_1b_3b_4} \lambda^{b_1} \lambda^{b_2} \lambda^{b_3} \lambda^{b_4}  \\
C_5 \!\!&=& \!\!\!\!\!\!\sum_{\stackrel{\scriptstyle{{a_1,a_2}}}{b_1,b_2,b_3,b_4,b_5}}
\!\!\!\!\!\!d_{a_1b_1b_2}d_{a_1a_2b_3}d_{a_2a_4b_5}  
\lambda^{b_1} \lambda^{b_2} \cdots \lambda^{b_5}\\
C_6 \!\!&=& \!\!\!\!\!\!\sum_{\stackrel{\scriptstyle{a_1,a_2,a_{3}}}{b_1,b_2,\dots,b_6}}
\!\!\!\!\!\!d_{a_1b_1b_2}d_{a_1a_2b_3}d_{a_2a_3b_4} 
d_{a_3b_5b_6} \nonumber \\
&&\times \lambda^{b_1} \lambda^{b_2} \cdots \lambda^{b_6}.
\eea
Of course the ones that are immediately interesting are $C_2, 
C_3,C_4,C_6,C_9$ for the purposes of 
embedding 2 qubits into a 4-state system, a 2-state and 3-state 
system into a 6-state system and the embedding of a two 3-state 
systems into a 9-state system.  These are useful for examining 
quantum control for two-qubits and entanglement issues for a 
two-qubits, a qubit and a qutrit, and two qutrits.  

The above relations can be expressed in terms of adjoint vectors and 
particular products.  We introduce this notation here since it 
has its own manipulation rules that make it easier to calculate 
quantities of interest.  Note also that since the $f_{abc}$ and 
$d_{abc}$ tensors are obtained by taking traces of products of 
elements with anticommutators and commutators respectively, 
they are 
easily calculated by analytic methods on a symbolic manipulation 
program such as {\texttt{MATHEMATICA}}.  These relations are 
$$
f_{abc} = \mbox{Tr}\left([\lambda_a,\lambda_b]\lambda_c\right),
$$
and 
$$
d_{abc} = \mbox{Tr}\left(\{\lambda_a,\lambda_b\}\lambda_c\right).
$$
The difference between upper and lower indices is not 
important if we are considering $SU(n)$.


\section{Trace Formulas}

\label{traceformulas}


\subsection{Symmetric Traces of Basis Elements}

Here the first few examples of the trace formulas have been given.  
\begin{widetext}
\bea
\!\!\!\!\tr(\lambda_i\lambda_j) &=& 2\delta_{ij} \;\;\;\;\;\;\;\;\;\;\;\\ 
\trs(\lambda_i\lambda_j\lambda_k ) &=& 2d_{ijk}\\ 
\trs(\lambda_i\lambda_j\lambda_k\lambda_l ) &=& \frac{4}{N}\delta_{ij}\delta_{kl} 
+ 2d_{ijm}d_{mkl} 
\eea
\bea
\trs(\lambda_i\lambda_j\lambda_k\lambda_l\lambda_q ) &=& 
	\frac{4}{N} (\delta_{ij}d_{klq} +\delta_{kl}d_{ijq}) + 2d_{ijm}d_{kln}d_{mnq}\\
\trs(\lambda_i\lambda_j\lambda_k\lambda_l\lambda_q \lambda_s) &=& 
	\frac{2^3}{N^2}\delta_{ij}\delta_{kl}\delta_{qs} \nonumber \\
	&&+ \frac{4}{N}(d_{ijm}d_{klm}\delta_{qs}  
	+ d_{ijm}d_{qsm}\delta_{kl} + d_{klm}d_{qsm}\delta_{ij} ) \nonumber \\
	&& + 2d_{ijm}d_{kln}d_{qst}d_{mnt} 
\eea
\bea
\trs(\lambda_i\lambda_j\lambda_k\lambda_l\lambda_q \lambda_s \lambda_u) &=& 
	   \frac{2^3}{N^2}(\delta_{ij}\delta_{kl}d_{qsu} 
	   + \delta_{ij}\delta_{qs}d_{klu}
	   + \delta_{qs}\delta_{kl}d_{iju}) \nonumber \\
	&& + \frac{2^2}{N}(\delta_{qs}d_{ijm}d_{kln}d_{mnu}
	   + \delta_{ij}d_{klm}d_{qsn}d_{mnu}
	   + \delta_{kl}d_{ijm}d_{qsn}d_{mnu}) \nonumber  \\
	&& + \frac{2^2}{N}d_{qsu}d_{ijm}d_{klm} \nonumber \\
	&& + 2d_{ijm}d_{kln}d_{qst}d_{mnt} 
\eea
\bea
\trs(\lambda_i\lambda_j\lambda_k\lambda_l\lambda_q \lambda_s \lambda_u \lambda_w) &=& 
	\frac{2^4}{N^3}\delta_{ij}\delta_{kl}\delta_{qs}\delta_{uw}\nonumber \\ 
	&&+\frac{2^3}{N^2}(\delta_{ij}\delta_{kl}d_{qst}d_{uwt}
	+\delta_{ij}\delta_{qs}d_{kln}d_{uwn}
	+\delta_{ij}\delta_{uw}d_{qsn}d_{kln} \nonumber \\
	&&\phantom{N^2+}\;\;+\delta_{kl}\delta_{qs}d_{ijn}d_{uwn}
	+\delta_{kl}\delta_{uw}d_{ijn}d_{qsn}
	+\delta_{uw}\delta_{qs}d_{kln}d_{ijn}) \nonumber \\
	&&+\frac{2^2}{N}(\delta_{ij}d_{klm}d_{qst} d_{uwv} d_{tvm}
	+\delta_{kl}d_{ijm}d_{qst} d_{uwv} d_{tvm} \nonumber \\
	&&\phantom{N^2+}\;\;+\delta_{qs}d_{ijm}d_{kln} d_{mnv} d_{uwv}
	+\delta_{uw}d_{ijm}d_{kln} d_{mnv} d_{qsv}) \nonumber \\
	&& +2d_{ijm} d_{kln} d_{mnp} d_{qst} d_{uwv} d_{tvp} 
\eea
\bea
\trs(\lambda_i\lambda_j\lambda_k\lambda_l\lambda_q \lambda_s \lambda_u 
		\lambda_w \lambda_y ) &=&
	\frac{2^4}{N^3}(\delta_{ij}\delta_{kl}\delta_{qs}d_{uyw}
	+\delta_{ij}\delta_{kl}\delta_{uw}d_{qsy}
	+\delta_{ij}\delta_{qs}\delta_{uw}d_{kly}
	+\delta_{kl}\delta_{qs}\delta_{uw}d_{ijy}) \nonumber \\
	&&+\frac{2^3}{N^2}(\delta_{ij}\delta_{kl}d_{qst}d_{uvw}d_{tvy}
	+\delta_{qs}\delta_{uw}d_{ijm}d_{kln}d_{mny} \nonumber \\
	&&\phantom{N^2+}+\delta_{ij}\delta_{qs}d_{kln}d_{uvw}d_{nvy}
	+\delta_{ij}\delta_{uw}d_{kln}d_{qst}d_{nty}) \nonumber \\
	&&+\frac{2^3}{N^2}(\delta_{ij}d_{kly}d_{qst}d_{uwt}
	+\delta_{kl}d_{ijy}d_{qst}d_{uwt} \nonumber \\
	&&\phantom{N^2+}+\delta_{qs}d_{ijm}d_{klm}d_{uwy} 
	+\delta_{uw}d_{ijm}d_{klm}d_{qsy}) \nonumber \\
	&&+\frac{2^2}{N}(\delta_{ij}d_{kln}d_{qst}d_{uvw}d_{tvx}d_{nxy}
	+\delta_{kl}d_{ijm}d_{qst}d_{uvw}d_{tvx}d_{mxy}\nonumber \\
	&&\phantom{N^2+}+\delta_{qs}d_{ijm}d_{kln}d_{mnp}d_{uwv}d_{pvy}
	+\delta_{uw}d_{ijm}d_{kln}d_{mnp}d_{qst}d_{pty})\nonumber \\
	&&+2 d_{ijm}d_{kln}d_{mnp}d_{qst}d_{uwv}d_{tvx}d_{pxy}
\eea


\subsection{Symmetric Traces for the Density Operator}

For the density operator these translate to (again only the first four are given):
\beq
\tr(\rho^2) \;=\; \frac{1}{N}\big[1 + (N-1)\nv\cdot\nv\big] \hspace{3in}
\eeq
\beq
\tr(\rho^3) \;=\;  \frac{1}{N^2}\big[1+3(N-1)\nv\cdot\nv+(N-1)(N-2) 
		(\nv\star\nv)\cdot \nv\big] \hspace{.85in}
\eeq
\bea
\tr(\rho^4) &=& \frac{1}{N^3}\big[1+6(N-1)\ndn \;\;\;\;\;\;\;\;\;\;\;\;\;\;\;\;\;\nonumber \\
	     && + 4(N-1)(N-2)(\nsn)\cdot\nv 
		 + (N-1)^2(\nv\cdot\nv)^2 \hspace{.05in}\nonumber \\
	     && \phantom{N^3}\;\;+ (N-1)(N-2)^2 (\nv\star\nv)\cdot(\nv\star\nv)\big] 
\eea
\bea
\tr(\rho^5) &=& \frac{1}{N^4}\big[1+10(N-1)\ndn + 10(N-1)(N-2)\nsn\cdot\nv \hspace{1.25in}
	\nonumber\\
	    &&\phantom{N^2+}+5(N-1)^2(\ndn)^2
		+5(N-1)(N-2)^2(\nsn)\cdot(\nsn) \nonumber \\
	    &&\phantom{N^2+}+2(N-1)^2(N-2)(\ndn)(\nsn\cdot\nv) \nonumber \\
	    &&\phantom{N^2+}+(N-1)(N-2)^3\nsn\star\nsn\cdot\nv\big]
\eea
\bea
\tr(\rho^6) &=& \frac{1}{N^5}\big[1+15(N-1)\ndn + 20(N-1)(N-2)\nsn\cdot\nv \hspace{1.3in}
	\nonumber\\
	    &&\phantom{N^2+}+15(N-1)^2(\ndn)^2
		+15(N-1)(N-2)^2(\nsn)\cdot(\nsn) \nonumber \\
	    &&\phantom{N^2+}+12(N-1)^2(N-2)(\ndn)(\nsn\cdot\nv) \nonumber \\
	    &&\phantom{N^2+}+6(N-1)(N-2)^3(\nsn\star\nsn\cdot\nv)\nonumber \\
	    &&\phantom{N^2+}+(N-1)^3(\ndn)^3+3(N-1)^2(N-2)^2(\ndn)(\nsn\cdot\nsn)
	\nonumber \\
	    &&\phantom{N^2+}+(N-1)(N-2)^4(\nsn\star\nv)^2\big]
\eea
\bea
\tr(\rho^7) &=& \frac{1}{N^6}\big[1+21(N-1)\ndn 
		+ 35(N-1)(N-2)\nsn\cdot\nv \nonumber\\
	    &&\phantom{N^2+}+35(N-1)^2(\ndn)^2
		+35(N-1)(N-2)^2(\nsn)\cdot(\nsn) \nonumber \\
	    &&\phantom{N^2+}+42(N-1)^2(N-2)(\ndn)(\nsn\cdot\nv) \nonumber \\
	    &&\phantom{N^2+}+21(N-1)(N-2)^3(\nsn\star\nsn\cdot\nv)\nonumber \\
	    &&\phantom{N^2+}+7(N-1)^3(\ndn)^3+21(N-1)^2(N-2)^2(\ndn)(\nsn\cdot\nsn)
	\nonumber \\
	    &&\phantom{N^2+}+7(N-1)(N-2)^4(\nsn\star\nv)^2 \nonumber \\
	    &&\phantom{N^2+}+ 3(N-1)^3(N-2)(\ndn)^2(\nsn\cdot\nv) \nonumber \\
	    &&\phantom{N^2+}+ 3(N-1)^2(N-2)^3(\nsn\star\nsn\cdot\nv)(\ndn) \nonumber \\
	    &&\phantom{N^2+}+ (N-1)(N-2)^3(\nsn\cdot\nv)(\nsn\cdot\nsn) \nonumber \\
	    &&\phantom{N^2+}+ (N-1)(N-2)^5(\nsn\star\nsn\star\nsn\cdot\nv)\big]
\eea
\bea
\tr(\rho^8) &=& \frac{1}{N^7}\big[1+28(N-1)\ndn 
		+ 56(N-1)(N-2)\nsn\cdot\nv \nonumber\\
	    &&\phantom{N^2+}+70(N-1)^2(\ndn)^2
		+70(N-1)(N-2)^2(\nsn)\cdot(\nsn) \nonumber \\
	    &&\phantom{N^2+}+112(N-1)^2(N-2)(\ndn)(\nsn\cdot\nv) \nonumber \\
	    &&\phantom{N^2+}+56(N-1)(N-2)^3(\nsn\star\nsn\cdot\nv)\nonumber \\
	    &&\phantom{N^2+}+28(N-1)^3(\ndn)^3+84(N-1)^2(N-2)^2(\ndn)(\nsn\cdot\nsn)
	\nonumber \\
	    &&\phantom{N^2+}+28(N-1)(N-2)^4(\nsn\star\nv)^2 \nonumber \\
	    &&\phantom{N^2+}+24(N-1)^3(N-2)(\ndn)^2(\nsn\cdot\nv) \nonumber \\
	    &&\phantom{N^2+}+24(N-1)^2(N-2)^3(\nsn\star\nsn\cdot\nv)(\ndn) \nonumber \\
	    &&\phantom{N^2+}+ 8(N-1)(N-2)^3(\nsn\cdot\nv)(\nsn\cdot\nsn) \nonumber \\
	    &&\phantom{N^2+}+ 8(N-1)(N-2)^5(\nsn\star\nsn\star\nsn\cdot\nv) \nonumber \\
	    &&\phantom{N^2+}+ (N-1)^4(\ndn)^4 + 6(N-1)^3(N-2)^2(\ndn)^2(\nsn\cdot\nsn) 
	\nonumber \\
	    &&\phantom{N^2+}+ 4(N-1)^2(N-2)^4(\ndn)(\nsn\star\nsn\star\nv\cdot\nv) 
	\nonumber \\
	    &&\phantom{N^2+}+ (N-2)^6(\nsn\star\nsn\star\nsn\star\nv\cdot\nv) \big]
\eea
\bea
\tr(\rho^9) &=& \frac{1}{N^8}\big[1+36(N-1)\ndn 
		+ 84(N-1)(N-2)\nsn\cdot\nv \nonumber\\
	    &&\phantom{N^2+}+126(N-1)^2(\ndn)^2
		+126(N-1)(N-2)^2(\nsn)\cdot(\nsn) \nonumber \\
	    &&\phantom{N^2+}+252(N-1)^2(N-2)(\ndn)(\nsn\cdot\nv) \nonumber \\
	    &&\phantom{N^2+}+126(N-1)(N-2)^3(\nsn\star\nsn\cdot\nv)\nonumber \\
	    &&\phantom{N^2+}+84(N-1)^3(\ndn)^3+252(N-1)^2(N-2)^2(\ndn)(\nsn\cdot\nsn)
	\nonumber \\
	    &&\phantom{N^2+}+84(N-1)(N-2)^4(\nsn\star\nv)^2 \nonumber \\
	    &&\phantom{N^2+}+108(N-1)^3(N-2)(\ndn)^2(\nsn\cdot\nv) \nonumber \\
	    &&\phantom{N^2+}+108(N-1)^2(N-2)^3(\nsn\star\nsn\cdot\nv)(\ndn) \nonumber \\
	    &&\phantom{N^2+}+ 36(N-1)(N-2)^3(\nsn\cdot\nv)(\nsn\cdot\nsn) \nonumber \\
	    &&\phantom{N^2+}+ 36(N-1)(N-2)^5(\nsn\star\nsn\star\nsn\cdot\nv) \nonumber \\
	    &&\phantom{N^2+}+ 9(N-1)^4(\ndn)^4 + 54(N-1)^3(N-2)^2(\ndn)^2(\nsn\cdot\nsn) 
	\nonumber \\
	    &&\phantom{N^2+}+ 36(N-1)^2(N-2)^4(\ndn)(\nsn\star\nsn\star\nv\cdot\nv) 
	\nonumber \\
	    &&\phantom{N^2+}+ 9(N-2)^6(\nsn\star\nsn\star\nsn\star\nv\cdot\nv) \nonumber \\
	    &&\phantom{N^2+}+ 4(N-1)^4(N-2)(\ndn)^3(\nsn\cdot\nv)  \nonumber  \\
	    &&\phantom{N^2+}+ 6(N-1)^3(N-2)^3(\ndn)^2(\nsn\star\nsn\cdot\nv) \nonumber  \\
	    &&\phantom{N^2+}+ 4(N-1)^3(N-2)^3(\ndn)(\nsn\cdot\nv)(\nsn\cdot\nsn) 
	\nonumber  \\
	    &&\phantom{N^2+}+ 2(N-1)^2(N-2)^5(\nsn\cdot\nsn)(\nsn\star\nsn\cdot\nv) 
	\nonumber  \\
	    &&\phantom{N^2+}+ 4(N-1)^2(N-2)^5(\ndn)(\nsn\star\nsn\cdot\nsn\star\nv) 
	\nonumber  \\
	    &&\phantom{N^2+}+ (N-1)(N-2)^7(\nsn\star\nsn\star\nsn\star\nv\star\nv\cdot\nv) 
		\big]
\eea
\end{widetext}


\begin{acknowledgments}
M.S.B. would like to thank the following people for their helpful 
discussions: William Wootters and Robert Griffiths (at QCMC '02), 
Randy Scott, Sara Schneider and especially Luis Boya and E.C.G. Sudarshan.  
M.S.B. would also like to thank Sara Schneider for a critical reading 
of the manuscript.  This work was supported by DARPA-QuIST Grant 
No. F49620-01-1-0556.

After the completion of this work, a detailed, independent proof of 
the positivity conditions \cite{Kimura}, was kindly pointed 
out by Gen Kimura.  This includes independent derivation of the 
Equations (\ref{eq:s1-s4}) and a more thorough discussion of the 
regions of positivity for three-state systems.
\end{acknowledgments}
    



\end{document}